# Development of Volume Produced Negative Ion Source using a CCRF Discharge

Pawandeep Singh[1,2,(a)], Swati Dahiya[1,2], Avnish Pandey[3], Yashashri Patil[1], Shantanu Karkari[1,2]

[1]*Institute for Plasma Research, Bhat, Gandhinagar—382428, India*
[2]*Homi Bhabha National Institute, Training School Complex, Anushaktinagar, Mumbai-400094, India*
[3]*Department of Applied Science and Humanities, United College of Engineering and Research, Prayagraj—211009, India*

[a)]*Corresponding author: Singh.pawandeep67@gmail.com*

## Abstract

**This work shows the development of a volume-produced negative ion source that consists of annular parallel plates driven by a 13.56 MHz capacitively coupled radio frequency in a push-pull configuration. This source shows advantages in controlling plasma conditions by varying the pressure, power, and applied axial magnetic field. It is found that the push-pull configuration allows the plasma potential to remain in the range of 20–40 Volts. Conversely, the application of a magnetic field helps serves to augment the production of negative ions in the central hollow part of the annular plate. Further, a plausible explanation to the obtained experimental results is presented.**

**Index Terms—Electronegative plasma, Negative ion diagnostic, Negative ion source**

## I. Introduction

Over the past several decades, research on negative ion plasmas has remained at the centre of attention due to its overwhelming applications in micro-electronic industries [1–4], the generation of energetic neutral particle beams for plasma heating in fusion devices [5,6], and its promising application in plasma propulsion [7–9]. These plasmas possess distinct characteristics that set them apart from electropositive plasmas. The primary factor behind this phenomenon is that the mass and temperature of negative ions are, respectively, comparable to those of positive ions, whereas their charge is equivalent to that of electrons. Owing to the ambipolar field, negative ions generally stagnate inside the bulk plasma volume. This led to a reduced value of positive ion flux at the sheath boundary. Therefore, efficiency of extracting negative ions from the bulk plasma remains quite small [10]. In a discharge, quantifying negative ion parameters is quite important from both research perspective as well as for the development of negative ion source.

Negative ions can be generated within the plasma volume through a two-step dissociative attachment process [11,12]. In the first step, neutral particles get excited to the metastable stage by collision with the highly energetic electrons. These metastable atoms undergo an attachment process, with the low-energy electrons producing negative ions in the second step. These negative ions can undergo additional loss mechanisms. One of the prominent loss mechanisms involves the detachment of negative ions through interactions with high-energy electrons and metastable atoms [12].

Since both high- and low-energetic electrons are required for the generation of negative ions, this work is primarily motivated towards the development of a negative ion source that has control over the spatial temperature distribution. In this study, a parallel plate annular capacitive radio frequency source is proposed. In this source, with the application of an axial magnetic field, the electron temperature is found to be high between the annular electrodes and decreasing towards the central hollow region. The self-generated DC bias appears to have a definable range of plasma potential with respect to the ground. Further, along with the basic source characteristic, this source is found to exhibit a high value of negative ion production in the central hollow region. This work is organised as follows:

Section 2 provides specifics of the experimental setup and diagnostic procedures employed in the study. Section 3 presents the experimental findings and their subsequent analysis and interpretation. Section 4 encompasses the comprehensive summary and conclusion of the study.

## II. Experimental Setup and Diagnostic

### A. Compensated Langmuir probe

Fig.-1 illustrates the experimental setup consisting of two annular parallel plates, with a separation distance of 10 cm [13]. The plates are powered by a 1:1 isolation transformer operating at a frequency of 13.56 MHz to create a discharge using oxygen as a feedstock gas. The outer diameter (b) and inner diameter (a) of the annular plates are kept at 20 cm and 10 cm, respectively.

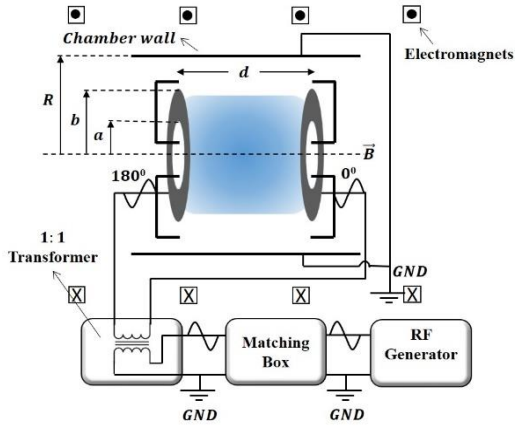

*Figure 1: Schematic of the experimental setup; a = 5 cm, b = 10 cm, R = 12 cm, d = 10 cm.*

The strength of the axial magnetic field is controlled using a set of four electromagnets that are spaced in a Helmholtz configuration. This arrangement results in a consistent magnetic field across a diameter of 24 centimetres, as depicted in Fig.-2.

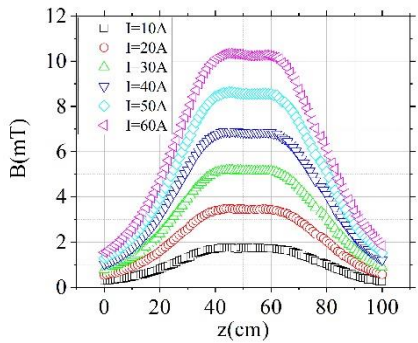

*Figure 2: For I=10-60A, X-axis is axial scale and Y-axis is magnetic field.*

The discharge is produced using an RF generator (RFG-600W coaxial power) in conjunction with a Pi-type (AMN-600R) automated matching unit. The power output of the matching unit is transmitted to the discharge electrodes through a ferrite transformer with a 1:1 turn ratio. The measurement of power supplied to the discharge electrodes is conducted using the Octive Poly 2.0 VI Probe.

### B. Compensated Langmuir probe

A Langmuir probe (LP) with a 5 mm length and 0.25 mm diameter made of tungsten has been used for the determination of plasma parameters. The LP is passively compensated using a series of self-resonating inductors. Fig.-3 shows the impedance that the probe offers to an applied peak-to-peak voltage of 5 Volts with varying frequency. Here, the frequency sweep is generated using a signal generator (Tektronix AGF3021C). The vertical lines drawn in the plot demarcate the impedance offered at the fundamental frequency (13.56 MHz) and its three harmonics.

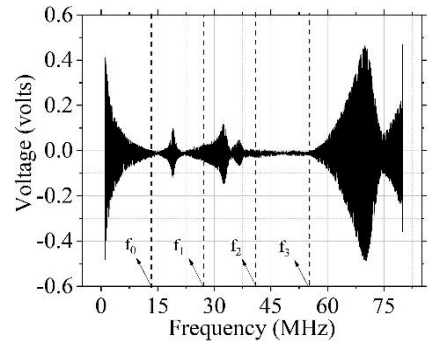

*Figure 3: Impedance offered by the compensated Langmuir probe to an applied voltage value of 5V peak-to-peak. $f_0$ is the fundamental 13.56 MHz frequency whereas $f_1$, $f_2$, and $f_3$ are the respective first, second, and third harmonic*

This probe is introduced in the mid-plane region between the annular plates. A train of triangular voltages is generated using a signal generator, which is subsequently amplified using a linear amplifier (WMA-300 Falc Systems) to obtain the probe characteristics. The frequency of these triangular voltages is kept at 11 Hz to avoid any capacitive pickup. The process of acquiring data from the Tektronix 3054C oscilloscope is performed using the LabVIEW software. The plasma parameters are derived from the acquired current-voltage characteristics of the Langmuir probe. The electron temperature is obtained by analysing the slope of the natural logarithm plot of electron current. The positive ion density is estimated by using the Orbital Motion Limited (OML) theory. Lastly, the plasma potential is determined by employing the first derivative method. The determination of the negative ion parameter associated with $O^-$ has been achieved by utilising the saturation current ratio (SCR) method, as negative atomic oxygen is identified as the prevailing species in the discharge. The details of the procedure to derive the negative ion parameter based on the saturation current ratio method can be found in our previous paper [14].

### III. Experimental results and Discussion

The experiment is carried out to characterise the capacitively coupled radio-frequency (CCRF) source for an applied power ranging from 10 to 100 watts and pressures from 0.1 Pa to 1 Pa using argon and oxygen as feedstock gases.

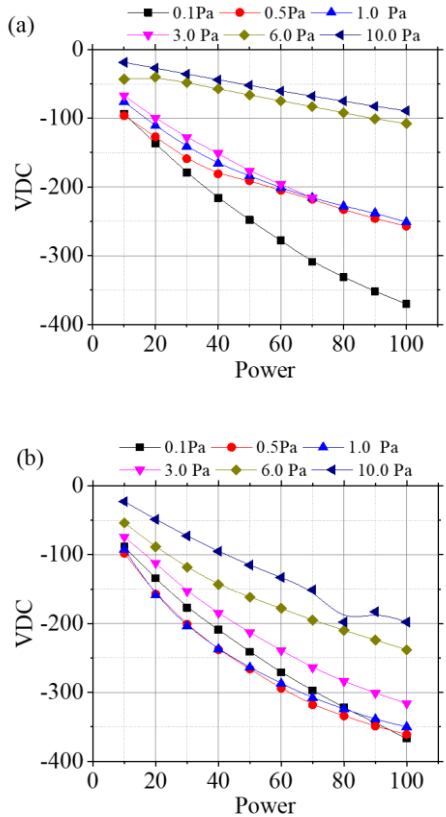

*Figure 4: Plot of self-bias voltage versus applied power in (a) argon discharge and (b) oxygen discharge for varying gas pressure.*

Fig.-4 shows the measurement of the induced self-DC bias voltage (VDC) as a function of applied power at the driven annular electrodes for both argon and oxygen. The experimental results indicate that the modulus of the VDC increases as the applied power increases, while it decreases with an increase in gas pressure at each specific point. The primary governing factors of this VDC are the electron temperature, plasma density, and collisions. The augmentation of power results in a proportional rise in plasma density when pressure is held constant, thereby elucidating the observed increase in VDC. Conversely, the reduction in VDC is concomitant with an augmentation in ion-neutral collisions corresponding to a constant value of applied power.

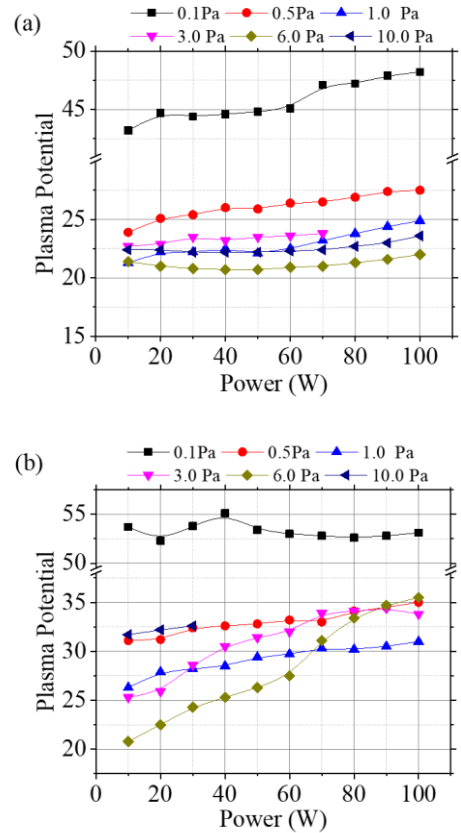

*Figure 5: Plot of plasma potential versus applied power in (a) argon discharge and (b) oxygen discharge for varying gas pressure.*

Fig.-5 displays the plasma potential values corresponding to various applied powers and pressures obtained using the compensated Langmuir probe. It has been observed that regardless of the increase in power, the plasma potential maintains a nearly constant value and exhibits a decrease with the increase in the background gas pressure from 0.1 to 1.0 Pa, which nearly saturates at high pressures (> 1.0 Pa). With the increase in an applied power, the RF voltage amplitude at the driven electrodes also increases. Since, plasma always tries to maintain a potential greater than any potential applied on the electrodes in its vicinity, the plasma potential would also increase with the increase in applied power in the present experiment. This would lead to a plasma potential in the range of 150 to 300 Volts. However, with the induced self-bias (VDC) on the driven electrodes, the centre of the RF voltage signal would shift negative depending upon the value of induced VDC. With this, the maximum voltage on the electrodes goes is of the order of 20 Volts with respect to the chamber and measurement ground. The observed uniform distribution of plasma potential with power and at constant pressure can be attributed to the synergistic influence of induced VDC.

Fig.-6 shows the plot of the radial profile of electron temperature and plasma density obtained using the compensated Langmuir probe. It is found that the electron temperature remains constant radially in the absence of a magnetic field. However, it results in a non-uniform electron temperature profile with the application of an axial magnetic field, where the central hollow region exhibits a lower electron temperature while the temperature progressively increases in an outward radial direction. This is because the region between the aligned electrodes (at r = 5 to 10 in Fig.-6) is primarily characterised by the prevalence of electrons with high energy, which in turn leads to the presence of hot electron populations with values ranging around 8 electron volts. However, low-energetic electrons exhibit diffusion across the magnetic field lines towards the central hollow region, giving rise to an observed range between 2 and 4 electron volts (eV).

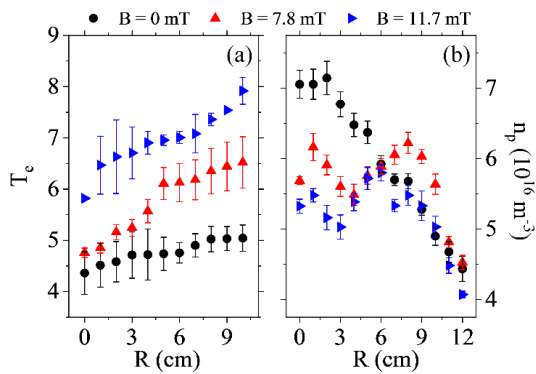

Figure 6: Plot of electron temperature and plasma density for a constant power of 50 W and working pressure of 0.33 Pa.

On the other hand, in the absence of a magnetic field, the plasma density (Fig.-6) follows the Bessel profile; however, with the application of an axial magnetic field, the plasma density at the central hollow region decreases and a double peak density profile appears. Since the plasma is primarily generated between the regions of the aligned electrodes (at r = 5 to 10), it diffuses to both sides (at r < 5 and >10). Because of the cylindrical geometry of the problem at hand, the diffusion inward and outward would be quite different. While for outward, it helps in reducing the density faster, for inward, it gives rise to further coagulation of plasma and hence increase in the plasma density, as evinced in Fig.-6.

The spatially resolved distribution of two-electron population within the discharge facilitates the effective generation of negative ions. This is obtained in the present experiment at r = 0 to 5 cm and r = 5 to 10 cm where the corresponding range of electron temperature is 5 to 7 eV and 6 to 8 eV, respectively. Fig.-7 shows the ratio of negative ion density to electron density, commonly referred to as plasma electronegativity ($\alpha = N_-/N_e$). The empirical observation reveals that the magnitude of plasma electronegativity exhibits an increase towards the central hollow region (at r = 0 to 5) as the magnetic field strength increases. Since, the confinement of energetic particles to the region of aligned electrodes (at r = 5 to 10) is enhanced by magnetic fields, this leads to an increased in the formation of metastable molecules. These metastable molecules subsequently undergo electron attachment in the central hollow region (at r = 0 to 5), resulting in the formation of negative ions. The plasma electronegativity decreases as one moves radially outward, as depicted in Fig.-7.

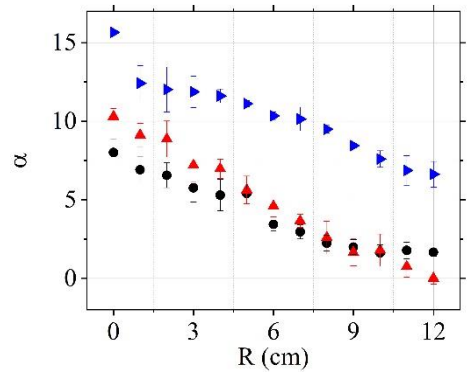

Figure 7: Plot of plasma electronegativity for a constant power of 50 W and working pressure of 0.33 Pa.

## IV. Summary and Conclusion

In this study, we have demonstrated a volume-produced negative ion source. This unique source configuration allows for spatially resolved high and low energetic electrons in the peripheral and central parts of the annular ring electrodes, respectively. It is further emphasised that this source allows the plasma potential to always remain in the range of 20 to 40 volts which is due to the self-induced DC bias to the discharge electrodes.

Both high- and low-energy electrons are often needed for the effective generation of volume-produced negative ions. The production of metastable atoms requires high-energy electrons, whereas the attachment process necessitates low-energy electrons. Since, this source provides the spatial resolution of two electron population, it is shown that a high value of plasma electronegativity can be found in the centre of the discharge which further increases with the application of magnetic field.

In general, this source offers a remarkable range of experimental parameters for conducting basic physics experiments. It is particularly valuable for testing theoretical models related to the discharge phenomenon and various diagnostic techniques.

## ACKNOWLEDGMENT

This work is supported by the Department of Atomic Energy, Government of India.